\documentclass[11pt,preprint]{aastex}

\begin{document}

\slugcomment{To appear in the Astrophysical Journal}

\title{Keck Interferometer observations of classical and weak line T Tauri stars}

\author{R.L. Akeson\altaffilmark{1}, A.F. Boden\altaffilmark{1}, J.D. Monnier\altaffilmark{2}, R. Millan-Gabet\altaffilmark{1}, C. Beichman\altaffilmark{1}, J. Beletic\altaffilmark{3}, N. Calvet\altaffilmark{4}, L. Hartmann\altaffilmark{4}, L. Hillenbrand\altaffilmark{5}, C. Koresko \altaffilmark{1}, A. Sargent\altaffilmark{5}, and A. Tannirkulam\altaffilmark{2} }

\altaffiltext{1}{Michelson Science Center, California Institute of Technology
, MS 100-22, Pasadena, CA, 91125}
\altaffiltext{2}{Univ. of Michigan, 941 Dennison Bldg, Ann Arbor, MI, 48109}
\altaffiltext{3}{W.M.Keck Observatory, California Association for Research in Astronomy, 65-1120 Mamalahoa Hwy., Kamuela, HI 96743}
\altaffiltext{4}{Smithsonian Astrophysical Observatory, Mail Stop 42, Cambridge, MA 02138}
\altaffiltext{5}{Dept of Astronomy, California Institute of Technology, MS 105-24, Pasadena, CA, 91125}

\begin{abstract}
We present observations of the T Tauri stars BP~Tau, DG~Tau, DI~Tau,
GM~Aur, LkCa~15, RW~Aur and V830~Tau, using long baseline infrared
interferometry at K band (2.2 $\mu$m) from the Keck Interferometer.
The target sources have a range of mass accretion rates and excess
near-infrared emission.  The interferometer is most sensitive to
extended emission on characteristic size scales of 1 to 5
millarcseconds.  All sources show evidence for resolved K band
emission on these scales, although a few of the sources are marginally
consistent with being unresolved.  We calculate the infrared excess
based on fitting stellar photosphere models to the optical photometry
and estimate the physical size of the emission region using simple
geometric models for the sources with a significant infrared excess.
Assuming that the K band resolved emission traces the inner edge of
the dust disk, we compare the measured characteristic sizes to
predicted dust sublimation radii and find that the models require a
range of dust sublimation temperatures and possibly optical depths
within the inner rim to match the measured radii.

\end{abstract}


\section{Introduction}

Circumstellar disks are an established component of the star formation
paradigm.  The circumstellar disk in the T Tauri stage can have several
attributes: a large, cold disk as traced by millimeter emission, a
substantial inner disk revealed by excess infrared emission and
accretion onto the central star measured by ultraviolet excess or
H$\alpha$ emission.  Young stellar objects are often classified by their
infrared spectral index \citep[Class I, II, III;][]{lad84} and by
their H$\alpha$ line-width (classical T Tauris with $>10$ \AA\ and
weak-line T Tauris with $<$ 10 \AA).  These classifications may
represent evolutionary stages, with Class I sources considered the
youngest with more massive disks and higher accretion rates,
although some recent work has questioned the age-evolution nature of
this sequence \citep{whi04}.  Accretion onto the central star from the
disk is dependent on and influences the inner disk ($<$1 AU)
properties.  Determination of the temperature and density structure of
the inner disk is important for understanding the mechanisms which
drive outflows and for establishing the initial conditions for planet
formation.  By studying disks around stars with a range of stellar and
disk properties we can begin to characterize the range of disk
properties and their relation to stellar properties.

Until the advent of infrared interferometry, the characteristics of
the inner disk were determined through spatially-unresolved
photometric or spectroscopic observations.  Near-infrared
interferometry is ideal for studying the inner disk, as these
observations are sensitive to hot material near the star itself and
given the milliarcsecond (mas) resolution capability of the current
generation of interferometers, these observations can in many cases
spatially resolve the emission from the inner disk.  Previous studies
\citep{mal98,mil99,ake00,mil01,tut01} have resolved the emission
around Herbig Ae/Be stars, FU Ori objects and T Tauri stars.  The
initial results for T Tauri and Herbig stars found characteristic
sizes larger than expected from geometrically flat accretion disks with
power-law temperature and density distributions
\citep{mil99,ake00,mil01}, while more recent observations of some Herbig
Be stars \citep{eis04} have found some objects which are consistent
with accretion disk predictions.  With the arrival of large aperture
interferometers, this technique can be applied to a much larger number
of objects covering a larger range of stellar and disk properties and
initial results have already appeared \citep{col03,lei04,eis05,mon05}.

Here we present Keck Interferometer observations of seven classical
and weak-line T Tauri stars with a range of inner disk and accretion
diagnostic observational signatures.  In \S \ref{source}, we describe
the source selection and characteristics of the sample and in \S,
\ref{observations} the infrared interferometry and photometry
observations.  Simple geometric models are fit to the data in \S
\ref{models} and the results are discussed in \S \ref{discussion}.

\section{Sources}
\label{source}

Candidate sources were selected with a
range of disk properties, as traced by the infrared excess and the
H$\alpha$ line width, and which could be tracked by the Keck
Interferometer in both the visible and infrared bands.  All sources
are in the Taurus-Aurigae cloud and we assume a
distance of 140~pc.  Given the visible sensitivity
requirement, the sample is biased against edge-on sources, because at
high inclinations the outer disk obscures the central region where the
visible light originates.  The source list with stellar and spectral
properties is given in Table \ref{table:source}.  For some of these
sources, a wide range of luminosities are estimated in the literature.
For consistency, the given luminosities are from a single
work where possible \citep{whi01} and from spectroscopic determinations
\citep{whi04} for the other sources.

\begin{table}[h!]
\begin{center}
\footnotesize
\begin{tabular}{lclllllllll} \hline
Source & Sp Type & Teff & L$_{\star}$(L$_{\odot}$) & M$_{\star}$(M$_{\odot}$)& R$_{\star}$(R$_{\odot}$)\tablenotemark{a}  & $\dot{M} (M_{\odot}/yr)$ &
Vmag\tablenotemark{b} & Kmag & F$_{\rm 1 mm}$(mJy) & Ref. \\ \hline 
BP Tau & K7 & 4060 & 0.83 & 0.77 & 1.9 & $1.3 \times 10^{-8}$ & 12.2 & 7.7 & 37 & 1,2,4 \\ 
DG Tau & K3 & 4775 & 3.62 & 2.2 & 2.8 & $7.4 \times 10^{-7}$ & 12.4 & 7.0 & 443 & 2,3,4
\\ 
DI Tau & M0 & 3850 & 0.62 & 0.4 & 1.79 & -- & 12.8 & 8.4 & $<$35 & 1,4 \\ 
GM Aur & K3 & 4730 & 1.01 & 1.22 & 1.5 & $6.6 \times 10^{-9}$ & 12.0 & 8.3 & 253 & 1,2,4 \\
LkCa15 & K5 & 4350 & 0.74 & 1.05 & 1.5 & $1.5 \times 10^{-9}$ & 12.4 &
8.2 & 167 & 1,2,5 \\ 
RW Aur & K2 & 4955 & 1.7 & 1.34 & 1.7 & $3.1 \times 10^{-8}$ & 10.5 & 7.0 & 421 & 2,5 \\ 
V830 Tau & K7 & 4060 & 0.79 & 0.77 & 1.83 & $<1.1 \times 10^{-9}$ & 12.2 & 8.4 & $<$9 & 1,2,5 \\ \hline
\end{tabular}
\caption{Stellar parameters for the observed sources.  
\label{table:source}}
\tablerefs{1) \citet{ken95} 2) \citet{whi01} 3) \citet{whi04} 4) \citet{bec90}
5) \citet{ost95}}
\tablenotetext{a}{Stellar radius derived from the stellar luminosity and effective temperature}
\tablenotetext{b}{Average V magnitude from \citet{ken95}}
\end{center}
\end{table}

\section{\bf Observations}
\label{observations}

\subsection{Infrared interferometry}
\label{KIobs}

Interferometry data were taken at the Keck Interferometer (KI), a
direct-detection infrared interferometer which connects the two
10-meter Keck telescopes.  The system includes adaptive optics for
each telescope with wave-front sensing in the visible, angle tracking
operating at J and H bands and fringe tracking at K band.  The fringe
tracking camera has a 50 mas FWHM field of view with a Gaussian
acceptance pattern.  KI is described in detail by \citet{col02} and
references therein.

Observations were taken from October 2002 to January 2004, over a
total of five nights.  The DG~Tau data were previously presented in
the analysis of \citet{col03}.  The data presented here are from the
wideband channel ($\lambda_{\rm center}$=2.14~$\mu$m and $\Delta
\lambda \sim$~0.3$\mu$m).  Observations consisted of a series of
interleaved integrations on the source and several calibrators.  The
sequence of calibrations and the standard KI data analysis method, in
which the visibility is corrected for the measured flux ratio of the
two arms, are described by \citet{col03}.  The data are presented as
the visibility amplitude squared,
normalized such that an unresolved object has V$^2=1.0$.

The system visibility, the instrumental response to a point source, is
measured using the calibrator stars listed in Table \ref{calibrators}.
Ideally, calibrator stars would match the target star at R (adaptive
optics), J (angle tracking) and K (fringe tracking) bands.  However,
given the colors of young stellar objects, these criteria cannot be
simultaneously met by standard stars and in choosing calibrators, the
angular distance and K band magnitude were given preference in the
matching.  Calibrators were shared between targets whenever possible
to optimize the observing efficiency.  The calibrator angular sizes
were estimated by modeling archival photometry from SIMBAD (references
in Table \ref{calibrators}) and the Two Micron All Sky Survey (2MASS).
All of the calibrators have angular diameters less than 0.3 mas and
thus are unresolved by the interferometer.  The calibrator angular
size uncertainties were set to 0.1 mas.  Source and calibrator data
were corrected for biases using sky calibrations as described by
\citet{col99} and averaged into blocks of 5 seconds each.  The data
were calibrated by dividing the measured visibility for each target by
the system visibility obtained from interpolating the bracketing
calibrator measurements, using the standard calibration package from
the Michelson Science Center.  The calibrated data points for the
target source are the average of the 15 to 25 5-second blocks in each
integration, with an uncertainty given by the quadrature of the
internal scatter and the uncertainty in the calibrator size.  As our
calibrators were unresolved, the size uncertainties have no
significant effect on the target calibrated visibilities.  The average
calibrated squared visibility, uncertainties and projected baseline
spatial frequency components ($u,v$) for each source are given in
Table \ref{table:KIdata}.

Given the limited amount of data collected, interpretation of the
results depends critically on determining the correct uncertainties in
the measurements, particularly for those objects which are marginally
resolved.  The visibility data presented here were taken during the
commissioning phase of the KI, when many system tests were also
undertaken.  The accuracy of the KI visibility mode was determined
through observations of binary targets with known orbits\footnote{see
http://msc.caltech.edu/KISupport/index.html for details} and it was
established that during the commissioning phase the systematic
uncertainty in V$^2$ for sources at this brightness level was 0.05.
Therefore, we add a systematic uncertainty of 0.05 in quadrature with
the measurement error.  Although any magnitude dependence of the
fringe tracker is not yet definitively established, each target source
is bracketed in K magnitude by its calibrators, which will minimize
any dependencies of the calibrated visibility on magnitude.

\begin{table}[h!]
\begin{center}
\small
\begin{tabular}{lllll} \hline
Calibrator & Ang. size & V & K & Sources \\ 
& (mas) \\ \hline
HD 282230 & 0.16$\pm 0.1$ & 9.1 & 8.6 & BP~Tau, DG~Tau, DI~Tau, GM~Aur, LkCa~15, RW~Aur, V830~Tau \\
HD 29050\tablenotemark{a} & 0.18$\pm 0.1$ & 8.9 & 6.8 & DG~Tau \\
HD 283668\tablenotemark{b} & 0.19$\pm 0.1$ & 9.4 & 7.0 & BP~Tau, DG~Tau, DI~Tau, LkCa~15, V830~Tau \\
HD 29334 & 0.12$\pm 0.1$ & 9.0 & 8.6 & DI~Tau, GM~Aur, LkCa~15, V830~Tau \\
HD 251383 & 0.17$\pm 0.1$ & 9.4 & 7.2 & GM~Aur \\
HD 36724 & 0.22$\pm 0.1$ & 7.6 & 6.3 & RW~Aur \\ \hline
\end{tabular}
\caption{Calibrator stars
\label{calibrators}}
\tablenotetext{a}{SIMBAD spectral type produced bad fit, used G8V template.}
\tablenotetext{b}{SIMBAD spectral type incorrect, used early K dwarf template.}
\tablerefs{Optical photometry references: Priser, J.~B.\ 1966, \pasp, 78, 
474; Rydgren, A.~E. et al.,
1984, \aj, 89, 
1015; Schuster, W.~J., \& 
Nissen, P.~E.\ 1988, \aaps, 73, 225; Slutskij, V.~E., 
Stalbovskij, O.~I., \& Shevchenko, V.~S.\ 1980, Soviet Astronomy Letters, 
6, 397 } 
\end{center}
\end{table}

\begin{table}[h!]
\begin{center}
\begin{tabular}{lllllll} \hline
Source & \# of ints. \tablenotemark{a} & Avg. V$^2$ & $\sigma_{\rm statistical}$ & $\sigma_{\rm stat. + systematic}$ & u (meters) & v (meters) \\ \hline
BP~Tau & 3 & 0.872 & 0.050 & 0.071 & 54.7 & 60.5 \\
DG~Tau & 5 & 0.383 & 0.011 & 0.051 & 42.0 & 73.4 \\
DI~Tau & 2 & 0.791 & 0.072 & 0.088 & 55.9 & 53.3 \\
GM~Aur & 4 & 0.862 & 0.031 & 0.059 & 55.1 & 47.4 \\
LkCa~15 & 2 & 0.895 & 0.018 & 0.053 & 55.2 & 54.1 \\
RW~Aur & 3 & 0.609 & 0.018 & 0.053 & 42.5 & 73.5 \\
V830~Tau & 2 & 0.879 & 0.042 & 0.065 & 51.3 & 64.0 \\ \hline
\end{tabular}
\caption{KI visibility data.
\label{table:KIdata}}
\tablenotetext{a}{An integration includes 125 seconds of fringe data and all necessary internal calibrations.}
\end{center}
\end{table}

\subsection{Photometry}

Visible and infrared photometry data for these sources were obtained
at the MDM Observatory 1.3m telescope and 
are listed in Table \ref{table:phot}.  The UBVRI photometry was based on
\citet{lan83} standards observed on 29 Nov 2004 using
Johnson UBVR and Kron-Cousins I filters in the 8k imager.  The JHK photometry
is based on bright infrared standards observed on 1 Dec 2004 
using Barr JHK filters in the TIFKAM instrument.

\begin{table}[h!]
\footnotesize
\begin{center}
\begin{tabular}{lllllllll} \hline
Source & U & B & V & R & I & J & H & K \\  \hline
BP~Tau & 13.20$\pm$0.15 & 13.36$\pm$0.03 & 12.32$\pm$0.04 & 11.45$\pm$0.04 & 10.60$\pm$0.04 & 9.10$\pm$0.10 & 8.37$\pm$0.10 & 7.90$\pm$0.10 \\
DG~Tau & 13.93$\pm$0.04 & 13.97$\pm$0.03 & 12.79$\pm$0.04 & 11.70$\pm$0.04 & 10.67$\pm$0.04 & --- & --- & --- \\
DI~Tau & 16.06$\pm$0.25 & 14.45$\pm$0.03 & 12.96$\pm$0.04 & 11.87$\pm$0.04 & 10.70$\pm$0.04 & --- & --- & --- \\
GM~Aur & 13.90$\pm$0.04 & 13.38$\pm$0.03 & 12.19$\pm$0.04 & 11.34$\pm$0.04 & 10.61$\pm$0.04 & --- & --- & --- \\
LkCa15 & 13.98$\pm$0.07 & 13.30$\pm$0.03 & 12.09$\pm$0.04 & 11.26$\pm$0.04 & 10.52$\pm$0.04 & --- & --- & --- \\
RW~Aur & 10.86$\pm$0.04 & 11.07$\pm$0.03 & 10.32$\pm$0.04 & 9.78$\pm$0.04 & 9.17$\pm$0.04 & 8.34$\pm$0.10 & 7.66$\pm$0.10 & 7.18$\pm$0.10 \\
V830~Tau & 14.66$\pm$0.15 & 13.52$\pm$0.03 & 12.21$\pm$0.04 & 11.26$\pm$0.04 & 10.44$\pm$0.04 & --- & --- & --- \\ \hline
\end{tabular}
\caption{Optical and infrared photometry from the MDM observations in magnitudes.
\label{table:phot}}
\end{center}
\end{table}

\section{Geometric models}
\label{models}

Given the limited spatial frequency ($u, v$) coverage of the data, we
use simple geometric models to characterize the size of the emission
region.  Only emission within the 50 mas field-of-view will contribute
to the measured visibilities.  We assume that the compact ($<$ 50 mas)
emission comprises the central star, which is unresolved at this
resolution, an incoherent contribution if an extended (i.e. completely
resolved) component is present and a partially resolved component.
Before the geometric models are fit to the data, we first consider the
possibility of stellar companions affecting the measured visibilities
(\S \ref{companions}), then attempt to quantify the possible
contribution from an extended component (\S \ref{extended}) and
finally quantify the stellar contribution through photometry fitting
(\S \ref{SED}).

\subsection{Known companions}
\label{companions}

We have searched the literature for known companions to our sources.
BP~Tau, DG~Tau, GM~Aur, LkCa~15 and V830~Tau have all been included in
multiplicity searches with speckle, HST imaging, and spectroscopy and
no companions were found \citep{ghe93,sar98,lei93,wal88}.  RW Aur and
DI Tau have known, but in our case, negligible companions,
which are discussed below.  Archival radial velocity data from the
Center for Astrophysics survey shows no significant variations for any
of the sources, except for RW~Aur A (Guillermo Torres, private
communication). Given the insignificant contributions of the known
companions, all targets are treated as single stars in the geometric
models.

RW Aur is a binary with a separation of 1\farcs4 and a K band flux
ratio of 4.3 \citep{whi01}.  The primary is well separated from the
secondary by the adaptive optics system at Keck and the secondary does
not contribute any flux because it lies well outside the fringe tracker
field-of-view.  \citet{gah99} reported spectroscopic variability in RW
Aur A and suggested that it may be a single-line spectroscopic binary
with a period of 2.8 days, although other causes of the variability
have not been ruled out.  If RW~Aur~A is a spectroscopic binary, the
measured radial velocity amplitude of 5.7 km/sec leads to a mass
function $(M_2)^3 \sin^3 i /(M_1 + M_2)^2$ of $5 \times 10^{-5}$
\citep{gah99}, which suggests a large mass ratio and thus a large
flux ratio at K.  It is therefore unlikely that the
resolved visibility is due to the possible companion.

DI Tau is a binary with a separation of 0\farcs12 and a K band flux
ratio of 8 \citep{ghe93}.  The secondary has not been detected in the
visible and the lower limit to the V band flux ratio is 21
\citep{sim96}.  The adaptive optics system will track within a few mas
of the primary and the incoherent flux contribution from the secondary
at K, including the tapering by the fringe tracker
field-of-view (FOV), is negligible ($2 \times 10^{-8}$).

\subsection{Extended components}
\label{extended}

Current imaging capabilities in the near infrared are not sufficient
to establish the distribution of flux on scales of several to 50 mas
(the KI FOV), which will contribute incoherently to the measured
visibility.  Possible sources of extended emission in these systems
include thermal or scattered emission from an envelope and scattered
light from the disk.  In this section, we discuss the limited
observational constraints and argue that, with one exception, there is
no evidence for substantial extended emission.  Any incoherent
contribution (which always decreases the measured visibility) not
accounted for results in an overestimate of the size of the partially
resolved component.

The observational technique which most directly addresses these size
scales is lunar occultation, which can probe emission down to several
mas.  In our sample, two objects, DI Tau and DG Tau have been observed
through lunar occultations.  Observations of DI Tau revealed the
secondary discussed in \S \ref{companions} \citep{che90}.  The
observations of DG Tau by \citet{lei91} revealed extended emission,
attributed to scattered light, which they fit with two components of
sizes 45 and 850 mas (FWHM) Gaussian.  The smaller component had a
flux ratio (extended/total) of 0.23.  \citet{che92} derived similar
results from independent observations.  We neglect the 850 mas
component as it contributes only 3\% of the total flux at K, but
include the 45 mas component in the modeling as an extended component
\citep[following~][]{col03}.  In a compilation of lunar
occultation observations, \citet{sim95} list 47 young stellar
systems in Taurus and Ophiuchus; of these only one (DG Tau)
is listed as having resolved emission and two are given as either
extended or binary, suggesting that substantial extended components
are not common on this size scale.

On larger observational scales, optical and infrared imaging can
reveal emission components which may extend into the central 50 mas
FOV around the star.  Using {\it Hubble Space Telescope} optical
imaging \citet{kri97} reported that LkCa 15 had no detectable
reflection luminosity and \citet{ghe97} found that RW Aur A was
consistent with a point source.

The issue of scattered light from the disk has been modeled in detail
for a few class II objects by \citet{ake05} who found that on much
larger scales (1 arcsec) the emission beyond 12 mas contributed less
than 6\% of the total flux at K.  These radiative equilibrium models
used the dust-size distribution derived through modeling of HH 30 IRS
and GM~Aur \citep{woo02a,sch03,ric03}.  Variations of the dust
properties with distance from the star, which would affect the
scattering properties, are not considered as this has not been studied
on the spatial scales probed by the KI observations.  For the sources
with a substantial infrared excess, an extended component at a level
of a few percent is within the error of the excess fraction and the
visibility measurement, particularly considering the smaller FOV for
KI.

All of these sources have been classified from their infrared SED as
class II ($d \log (\lambda F_{\lambda})/d \log \lambda \approx $ -2 to
0) except V830 Tau, which is class III ($d \log (\lambda F_{\lambda})/d
\log \lambda \approx $ -3) \citep{ken95}.  In general, Class
II sources are thought to have little or no envelope remaining \citep[see e.g.][]{mun00}.  For an envelope to contribute incoherently
in these measurements, it would have to have significant emission
(compared to the stellar photosphere and thermal disk emission)
on size scales less than 7 AU.

Given the lack of detected extended emission, except for DG Tau, and
the work on scattering from disks in similar sources, we assume for
all sources except DG Tau that there is no extended component. 
If the measured visibilities were due entirely to a scattered or
extended component, the flux ratio (extended/stellar) would
range from 0.07 to 0.62 for our sample.

\subsection{Spectral energy distribution modeling}
\label{SED}

Determining the stellar contribution at K-band is essential in
interpreting the visibility results.  To do this, we modeled optical
photometry from the literature and this work with stellar photosphere
spectral energy distribution (SED) templates based on the effective
temperatures given in Table \ref{table:source}.  The synthetic SEDs
were taken from the libraries created by Kurucz (2001) and
\citet{leu97}, using the solar abundance models with an effective
temperature matching that given in Table \ref{table:source} and for
surface gravity $\log g = 4.0$, appropriate for young pre-main
sequence stars.  The synthetic SEDs were integrated over the
appropriate passband for comparison to the photometry.  In general, we
found better fits at optical wavelengths using the Lejeune models, and chose
to standardize on that family of templates for the present analysis.
The Lejeune models have a correction function applied which yields
synthetic colors matching the empirical colors derived from
observations of main sequence stars \citep{leu97} and these models
better fit the optical spectral shape of our targets.  For a fixed
template/effective temperature, photosphere apparent size and
extinction ($A_V$) were used as free parameters in our modeling, and
to avoid biases from hot accretion luminosity and cooler circumstellar
material we used only the $V$, $R$, and $I$ photometry.

\begin{figure}[h!]
\begin{center}
\epsscale{0.9}
\plotone{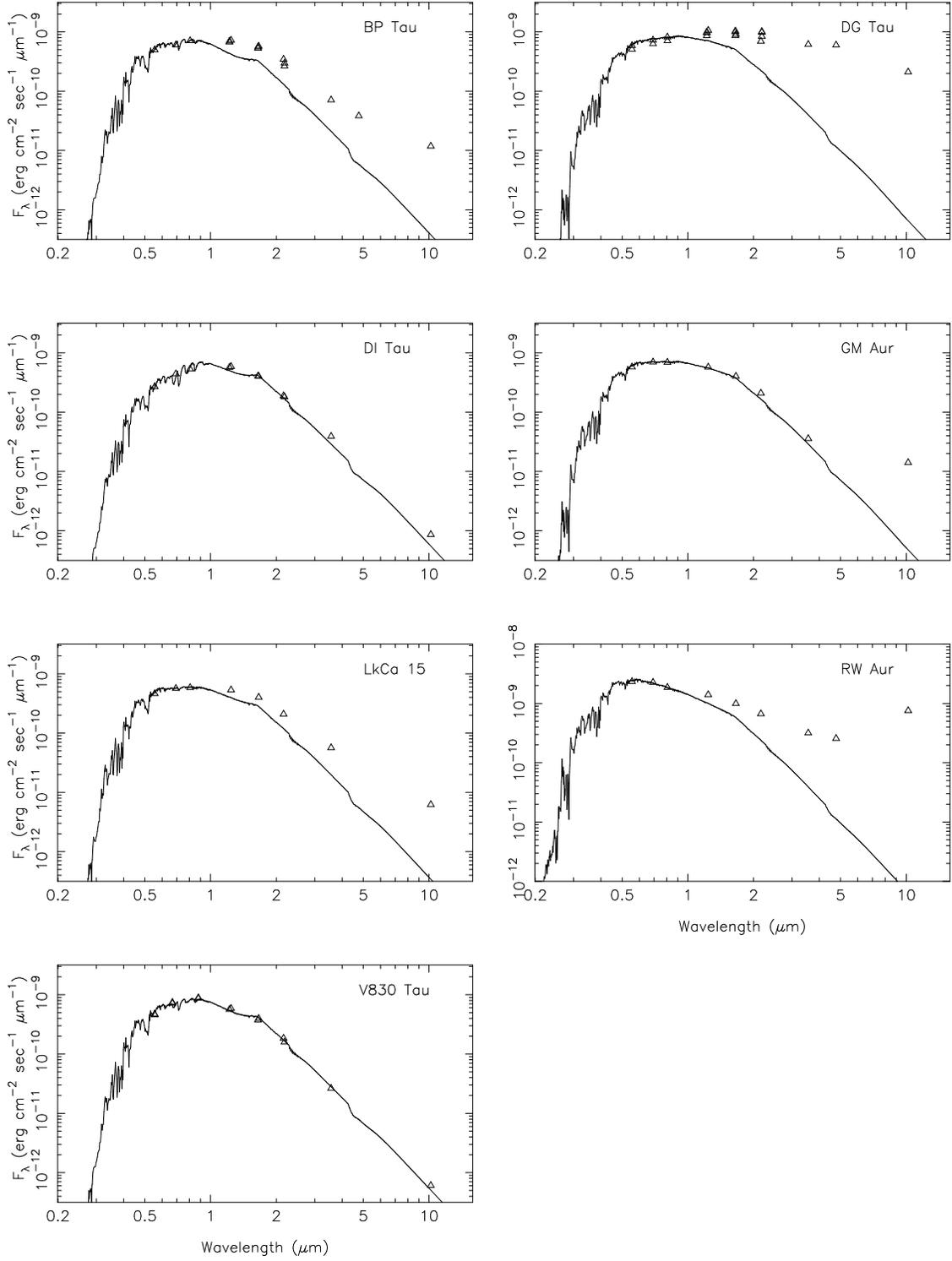}
\caption{Optical and infrared photometry (points) for the target
sources shown with the Lejeune stellar photosphere model, including
the correction for the derived extinction.
\label{fig:SED}}
\end{center}
\end{figure}

Results from the SED fitting are summarized in Table
\ref{table:SEDfit} and Figure \ref{fig:SED}.  The fractional K-band
excess is estimated by taking the K photometry measurement closest
in time to the interferometry data (in all cases this is either 2MASS
or from Table \ref{table:phot}).  Here we define the fractional excess
as the non-stellar emission divided by the total emission at K.
The uncertainty in the fractional
excess is taken to be the combination of the measurement uncertainty
and an estimate of the uncertainty in the stellar contribution at K.
We considered two possible components to the stellar contribution
uncertainty: variations in the optical photometry and therefore in the
fit extinction; and variations in the infrared photometry, as the
photometry and interferometry data are not contemporaneous.  For these
sources, the infrared variability is the dominant component and was
used to determine the stellar contribution uncertainty.  The infrared
variability was taken to be the larger of 1) the scatter in infrared
excess using K photometry fit to our stellar model or 2) the scatter
in K photometry as collected by \citet{ken95}.  Our estimated K-band
excess and uncertainty are given in Table \ref{table:SEDfit}.  Note that
GM Aur had relatively few measurements in our literature search, and
thus our variability estimate may be low.

\begin{table}[h!]
\begin{center}
\begin{tabular}{lcllll} \hline
Source & Model Teff (K) & Av & Fractional K excess & $\sigma$ K excess\tablenotemark{a} & Photometry ref \\ \hline
BP Tau & 4000 & 0.078 & 0.56 & 0.12 & 1,2,3,4 \\ 
DG Tau & 4750 & 1.57 & 0.67 & 0.19 & 1,2,3 \\
DI Tau & 3750 & 0.54 & 0.07 & 0.10 & 1,2,5 \\
GM Aur & 4750 & 1.21 & 0.12 & 0.02 & 1,2,3 \\
LkCa15 & 4250 & 0.46 & 0.43 & 0.04 & 1,2,6 \\
RW Aur & 5000 & 0.24 & 0.64 & 0.10 & 1,2,7 \\
V830 Tau & 4060 & 0.38 & 0.06 & 0.12 & 1,2,3,8 \\ \hline
\end{tabular}
\caption{Results of photometry fitting.  
\label{table:SEDfit}}
\tablerefs{1) this work 2) 2MASS 3) \citet{ryd83} 4) \citet{bou88} 
5) \citet{ryd81} 6) \citet{her86} 7) \citet{whi01} 8) \citet{mun83}
}
\tablenotetext{a}{The uncertainty includes both the measurement error and an estimate of the uncertainty in the stellar component.}
\end{center}
\end{table}

This method assumes that the VRI photometry is due entirely to
emission from the stellar photosphere.  Problems may arise at the
shorter wavelengths with contamination from accretion luminosity and
veiling and at the longer wavelengths by emission from the disk
itself.  DG~Tau is known to have significant veiling \citep{joh01} and
is highly variable.  We were unable to fit the entire photometry
collection on this source with the selected stellar template.  To
check our method of determining the infrared excess, we compared the
values in Table \ref{table:SEDfit} to measurements of the K band
veiling (here converted to fractional excess at K) where available.
Our determination of the K band excess does agree at the 1$\sigma$
level with veiling estimates from the literature, although for BP~Tau,
our estimate of 0.56 $\pm$ 0.12 is higher than all three veiling
estimates: 0.42 $\pm$ 0.05 \citep{joh01}, 0.44 $\pm$ 0.23
\citep{fol99} and 0.37 $\pm$ 0.17 \citep{muz03}.  The other sources
with veiling measurements in the literature are (our fractional excess
listed first): DG~Tau, 0.67 $\pm$ 0.19 vs. 0.59 $\pm$ 0.07
\citep{joh01}; GM~Aur, 0.12 $\pm$ 0.02 vs. 0.23 $\pm$ 0.21
\citep{fol99} and RW~Aur, 0.63 $\pm$ 0.10 vs. $>0.6$ \citep{fol99}.  A
smaller fractional excess would result in a larger fit radius for a
given visibility.

\subsection{Sources without a measured infrared excess}

The weak-line T Tauri sources DI~Tau and V830~Tau show no significant
excess in their SED with our modeling procedure (Figure
\ref{fig:SED}).  In our SED modeling, the K band excess for these two
sources is 0.07$\pm$0.10 and 0.06$\pm$0.12, respectively, consistent
with no near-infrared excess.  The lack of simultaneous infrared
photometry prevents us from a more accurate measurement of the
infrared excess.  Our fitting is consistent with a lack of excess
determined previously for these sources at both near-infrared and
mid-infrared wavelengths \citep{mey97,sta01,met04}.  Neither of these
sources has been detected at millimeter wavelengths (Table
\ref{table:source}) which probe cooler disk material.

The measured visibilities, 0.79$\pm$0.08 and 0.88$\pm$0.07, are within
3$\sigma$ of an unresolved point source.  For comparison, we
considered the source HD 283798 (RXJ 0441.8+2658) which was mistakenly
included as a calibrator.  This source was identified as pre-main
sequence by \citet{wic96} based on its x-ray flux and lithium line
width, but has no infrared excess.  The calibrated V$^2$ for HD~283798
is 0.98$\pm$0.06, consistent with an unresolved source, demonstrating
that not all observed sources are resolved.

One possible source of systematic error is that the science sources
are all dimmer than the calibrators (and HD 283798 analyzed above) in
the visible, where the adaptive optics (AO) wave-front sensing
operates.  The measured visibility will be reduced if the AO
correction is degraded and therefore the visibility may be a function
of the visible magnitude.  As part of the commissioning tests to
determine the system performance of the interferometer (see \S
\ref{KIobs}) attenuation was introduced in the AO system to
artificially dim known calibrators.  During one of these tests, the
same calibrator (HD 3765) was observed in two consecutive integrations
with and without AO attenuation.  The true apparent V magnitude of
HD 3765 is 7.4 and the attenuated apparent V mag was 11.9.  The
measured visibility, corrected for the ratio imbalance, was 0.65 $\pm$
0.08 with the AO attenuation and 0.67 $\pm$ 0.05 without the AO
attenuation.  Although our science targets are up to a magnitude
fainter than V=11.9, they also have a redder spectrum (V-K = 3.5 to
5.4) than HD 3765 (V-K=2.3) such that the I band magnitudes, still
inside the A0 wave-front sensor spectral sensitivity, were more
similar.  In this limited (few integration) comparison, there is no
substantial decrease in the visibility at the fainter V level,
suggesting any systematic is minimal.

There are several possible causes for the measured resolved visibilities
of DI Tau and V830 Tau: unknown companions within the 50 mas field of
view, a small fractional excess in the near-infrared which our
modeling procedure is not accurate enough to definitively measure, or 
an extended component due to scattered light, which, having the same
spectrum as the star, would not appear as an excess in our fitting
procedure.  As there is no photometric evidence for a disk around
these sources at any wavelength, we have not fit them with a disk
model as we do for the other sources.  However, we can quantify the
scattered flux or unknown companion possibilities given the measured
visibility.  For the unknown companion, we assume that the separation
is larger than the fringe spacing (as no radial velocity variations are
seen) such that the potential companion would contribute incoherently
to the visibility.  We also assume that any potential scattered light
is extended and therefore contributes incoherently.  The ratio of
scattered or companion flux to stellar flux would be 0.13 $\pm$ 0.007 for DI
Tau and 0.067 $\pm$ 0.003 for V830 Tau to account for the measured
visibilities.

\subsection{Sources with an infrared excess}

The sources BP~Tau, DG~Tau, GM~Aur, LkCa~15 and RW Aur all show a
significant near-infrared excess using our fitting method.  All of
these sources also have a mid-infrared excess (Figure \ref{fig:SED}) 
and have been detected
at 1~mm (Table \ref{table:source}).  These longer wavelength
measurements suggest the presence of an extended circumstellar disk
and in the case of BP Tau, DG Tau, GM Aur, and LkCa 15 the cooler,
outer disk has been mapped at millimeter wavelengths
\citep{kit96,sim00,qi03}.

For each of these sources, we use a model for the compact ($<$ 50 mas)
emission which consists of the central star, which is unresolved at
this resolution, an incoherent contribution from known scattered light
(DG Tau only) and a resolved component.  To estimate the size of the
resolved component, we use two simple geometric models, a uniform disk
and a ring.  In each case, the interferometry data is used to measure
the emission region size, after accounting for the stellar and
scattered components.  For the ring model, the width is determined by
the excess flux at K, for a blackbody temperature of 1500~K.  Given
the limited spatial frequency coverage of the observations, the data
do not constrain the disk inclination and face-on models are used.
This will underestimate the true size if the source emission has a
ring-like morphology and is substantially inclined from face-on, but
as stated earlier, our sample is biased against highly inclined
sources.  Estimates have been made for the inclination of some of
these sources based on other observations.  Using millimeter CO maps,
\citet{sim00} fit an inclination angle of 30$\pm$3\arcdeg\ for BP~Tau,
56$\pm$2\arcdeg\ for GM~Aur and 52$\pm$1\arcdeg\ for LkCa~15 and
\citet{qi03} derived an inclination angle of 57$\pm$5\arcdeg\ for
LkCa~15.  The measured radii for the two geometric models are given in
Table \ref{table:V2fit}.  The model uncertainty is given for both the
uncertainty in the data (scatter and 0.05 systematic component) and
for the uncertainty in the K band excess, $\sigma_K$.

\begin{table}[h!]
\begin{center}
\footnotesize
\begin{tabular}{lllll|llllll} \hline
 & \multicolumn{4}{c}{Uniform Disk} & \multicolumn{5}{c}{Ring} \\
Source & Radius & $\sigma_K$ & Radius & $\sigma_K$ & Radius & Width &  $\sigma_K$ & Radius & Width & $\sigma_K$  \\ 
& (mas) & (mas) & (AU) & (AU) & (mas) & (mas) & (mas) & (AU) & (AU) & (AU) \\ \hline 
BP Tau & 0.880$^{+ 0.295}_{-0.395}$ & 0.120 &  0.123$^{+ 0.041}_{-0.055}$ & 0.017 &  0.595$^{+ 0.210}_{ 0.295}$ & 0.043 & 0.090 &  0.083$^{+ 0.029}_{ 0.041}$ & 0.006 &  0.013 \\

DG Tau & 1.505$^{+ 0.335}_{-0.340}$ & 0.615 &  0.211$^{+ 0.047}_{-0.048}$ & 0.086 &  1.015$^{+ 0.225}_{ 0.240}$ & 0.060 & 0.415 &  0.142$^{+ 0.031}_{ 0.034}$ & 0.008 & 0.058 \\

GM Aur & 2.320$^{+ 1.065}_{-0.920}$ & 0.180 &  0.325$^{+ 0.149}_{-0.129}$ & 0.025 &  1.580$^{+ 0.585}_{ 0.605}$ & 0.003 & 0.240 &  0.221$^{+ 0.082}_{ 0.085}$ & 0.001 &  0.034 \\

LkCa 15 & 1.025$^{+ 0.345}_{-0.475}$ & 0.045 &  0.143$^{+ 0.048}_{-0.066}$ & 0.006 &  0.710$^{+ 0.240}_{ 0.340}$ & 0.020 & 0.035 &  0.099$^{+ 0.034}_{ 0.048}$ & 0.003 &  0.005 \\

RW Aur & 1.660$^{+ 0.205}_{-0.205}$ & 0.165 &  0.232$^{+ 0.029}_{-0.029}$ & 0.023 &  1.145$^{+ 0.140}_{ 0.140}$ & 0.015 & 0.115 &  0.160$^{+ 0.020}_{ 0.020}$ & 0.002 &  0.016 \\ \hline
\end{tabular}
\caption{Fit radii and uncertainties for the uniform disk and ring models.
$\sigma_K$ is the uncertainty due to the uncertainty in estimating the
K band excess.
\label{table:V2fit}}
\end{center}
\end{table}

\section{Discussion}
\label{discussion}

Previous studies of young stellar objects with infrared interferometry and disk
modeling have argued that the resolved component traces emission from
the inner edge of the disk \citep{mil01,tut01,ake02}.  On the theoretical
side, disk models have been proposed in which the gas component is
optically thin, and the infrared emission of the inner disk is
dominated by emission at the radius where the dust sublimates
\citep{nat01,dul01,muz03}.  In some high accretion cases, the emission from
the gas may contribute to the infrared emission and therefore to the
measured visibility as discussed by \citet{ake05}.  \citet{naj03}
observed fundamental CO emission from two stars in our sample and
derived inner radii for the gas emission of 0.043 AU for BP~Tau and
0.093 AU for LkCa~15.  The corresponding ring radii from our data are
0.083 AU (BP Tau) and 0.099 AU (LkCa~15).  As we have not accounted for
possible inclination effects, this is consistent with some gas being
present within the inner dust radius.

Several groups have recently worked on the dust sublimation
radius in models of circumstellar disks.  \citet{mon02} considered
the case of no backwarming of the grains (i.e. an optically thin
ring), which represents the lower limit to the dust sublimation radius.
\citet{dul01} developed a physical model for irradiated dust disks,
including heating within the disk,
and examined the vertical extent of the inner rim.  \citet{muz03} and
\citet{dal04} extended this model to include the heating due to the
accretion shocks.   
In each of these 3 models, the dust sublimation radius can be parameterized as
\begin{equation}
R_{dust} = f (\frac{L}{L_{\odot}})^{1/2} (\frac{T_{dust}}{1500K})^{-2} AU
\label{Rdust}
\end{equation}
where T$_{dust}$ is the dust sublimation temperature.  To calculate
the numerical factor $f$ we used the dust absorption efficiencies and
geometric factors given in each paper as appropriate for T Tauri
stars.  The values of $f$ are 0.047, 0.072 and 0.069 respectively for
\citet{mon02,dul01,muz03}.  A somewhat different approach was taken by
\citet{whit04} who used a varied grain size model and iteratively
solved for the dust sublimation radius in their Monte Carlo radiative
transfer disk models.  They parameterize R$_{dust}$ as a function of
the stellar radius and temperature, which we converted to luminosity
and temperature to compare to the previously discussed models.  The
luminosity dependence is the same, but the temperature exponent in
\citet{whit04} is -2.085 and there is a small dependence on the
stellar temperature (R$_{dust} \propto$ T$_{\star}^{0.085}$).  To
compare this model with the others, we use a single stellar
temperature (4500~K) but this introduces only a few percent error.  We
calculated R$_{dust}$ for each of these models; with the result that
the \citet{mon02} model delineates the lower bound for R$_{dust}$ as a
function of luminosity, while \citet{whit04} traces the upper bound.
We note that values of R$_{dust}$ predicted by the optically thick
\citet{dul01} and \citet{muz03} models are within 10\% of the
\citet{whit04} values for the luminosity range of these sources, and
we therefore use the \citet{mon02} (hereafter, optically thin) and
\citet{whit04} (hereafter, optically thick) models to represent the
range of possible R$_{dust}$.  In these models, the optically
thin and thick designations refer only to the properties of
the disk inner rim; in both models, only optically
thin material is present interior to the dust sublimation
radius.
Based on the results of the
\citet{mon05} sample of Herbig stars and the \citet{muz03} sample of T
Tauri stars, we use example dust sublimation temperatures of 1000 and
1500~K.

We compare our measured ring radii to R$_{\rm dust}$ using the
combined stellar and accretion luminosities following \citet{muz03}.
Accretion luminosities can be difficult to determine for T Tauri stars
and estimates for individual sources can have a wide range from
different methods.  For consistency, we use 
a derivation of the accretion luminosity from \citet{har98},
\begin{equation}
L_{\rm acc} = 0.8 G\dot{M}_{\rm acc}M_{\star}/R_{\star}
\end{equation}
where the numerical factor of 0.8 assumes dissipation of accretion energy
as material falls freely in along the stellar magnetosphere.
The mass accretion rate, stellar mass and stellar radius for each source are
taken from Table \ref{table:source}.
For DG Tau, which has the highest accretion rate in
the sample, we used the accretion luminosity of 2.8 L$_{\odot}$ from
\citet{whi04}.  For the other sources, the derived accretion
luminosities are 0.13, 0.12, 0.03 and 0.62 L$_{\odot}$ for BP~Tau,
GM~Aur, LkCa~15 and RW~Aur respectively.  We note that the stellar
parameters for DG Tau used here (taken from \citet{whi04} and based on
spectroscopy observations) have a significantly higher stellar
luminosity than the estimates (L$_{\star}$ = 1.4 L$_{\odot}$,
\citet{joh01} and L$_{\star}$ = 1.7 L$_{\odot}$, \citet{bec90}) used for the previous analysis of the DG~Tau data \citep{col03}.
Given the variations in luminosity determinations, we assign
uncertainties of 25\% for the stellar luminosities and 50\% for the
accretion luminosities.  The mass accretion rates in Table
\ref{table:source} agree to within these uncertainties with other
estimates where available \citep{gul98}.

\begin{figure}[h!]
\begin{center}
\epsscale{1.0}
\plotone{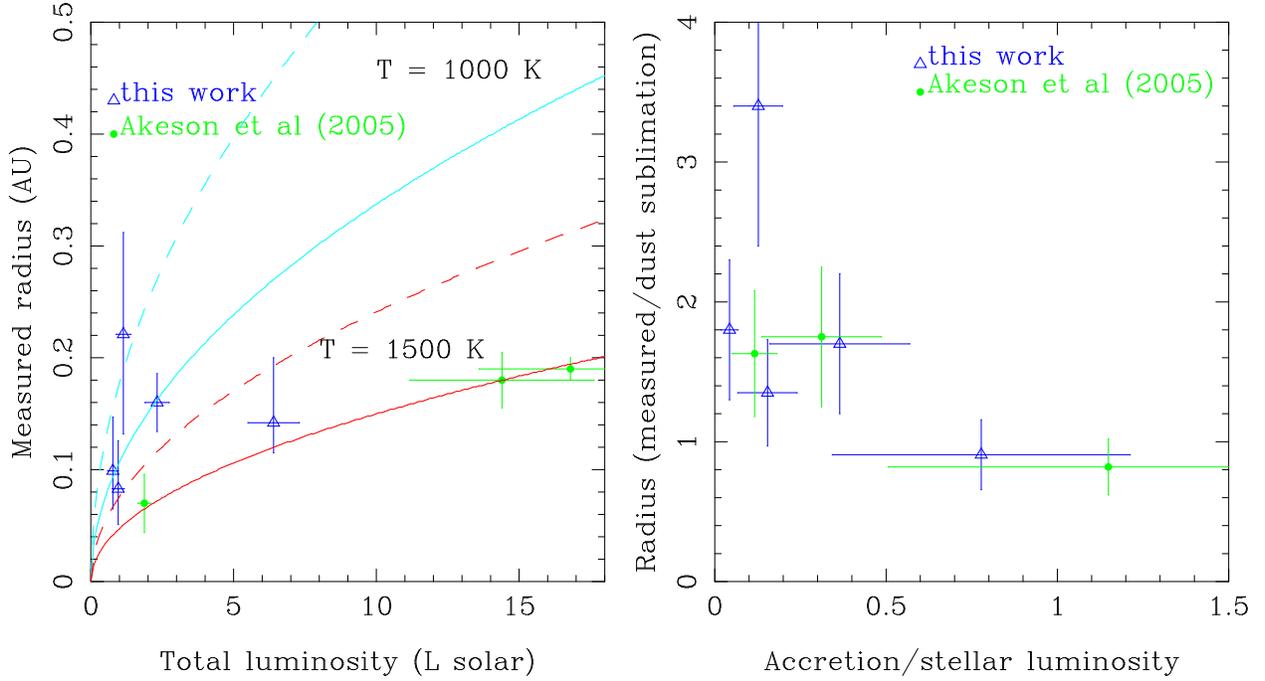}
\caption{Comparison of the measured radius and the dust sublimation
radius for 5 sources from this work (triangles) and 3 sources from
Akeson et al 2005 (filled circles).  The left panel shows the
measured ring radius versus the total (stellar and accretion)
luminosity.  The top two lines are the \citet{mon02} (solid line) and
\citet{whit04} (dashed line) models with a dust sublimation temperature of 1000~K and
the bottom two lines are the same models with a dust sublimation
temperature of 1500~K.  The right panel shows the ratio of the measured
radius to the dust sublimation radius as a function of ratio of
accretion and stellar luminosity.  In this case, the dust sublimation
radius is defined as the mean of the \citet{mon02} and \citet{whit04}
models with the error bars showing the spread between those two.
\label{fig:dust}}
\end{center}
\end{figure}

Figure \ref{fig:dust} (left panel) shows the measured ring radius
against the total (stellar and accretion) luminosity for the 5 KI
sources with infrared excesses and 3 additional T Tauri sources from
the PTI observations of \citet{ake05}.  The solid lines represent the
predicted dust sublimation radius from the optically thin
model using T$_{dust}$ of 1000 and 1500~K and the dotted lines
are the optically thick model at the same two temperatures.  

Before discussing the comparison, we note possible issues with some of
the individual sources.  For RW~Aur, other groups have calculated
higher stellar luminosities \citep[e.g. 3.5 L$_{\odot}$;][]{har95}
than used here.  The SED of GM~Aur (the source with the largest
measured radius) has led to models of large inner holes in the disk
\citep{ric03} and the resolved visibility could be due to scattering
off an inner rim at several AU.  However, these groups have calculated
lower extinctions and no infrared excess \citep{ric03,sch03} using
older infrared photometry \citep{coh79}, whereas the 2MASS photometry
used here is 0.3 magnitudes brighter in the near-infrared and we found
a small but significant infrared excess in our fitting procedure (although as
noted in \S \ref{SED}, our variability estimate may be low).  We
are also using a spectral temperature of 4730~K from \citet{whi01}
which is higher than the 4000~K used by previous studies.  If there are
substantial scattered light components in the central 50 mas (7 AU)
for sources other than DG~Tau then our radii will be overestimates of
the disk radius.  However, if the sources have non-face-on
inclinations (outer disk inclination estimates are 30\arcdeg\ to
60\arcdeg), then the measured radii are an underestimate of the disk
radius.  Note that for the PTI sources \citep{ake05}, three separate
baselines were used and these sources were fit with an inclined ring.

If the resolved visibilities do represent the inner disk radius, a
single inner disk model (optically thick vs. optically thin) and dust
sublimation temperature does not fit the sample here.  If GM~Aur is
discounted, then the optically thin model (Figure 2, solid lines) with
a range of T$_{dust}$ from 1000 to 1500~K can fit the sample, but the
optically thick inner rim does not match all the objects for this
range of T$_{dust}$ and would require substantially higher T$_{dust}$
to fit the sample.  Infrared interferometric sizes have been measured
for a larger number of Herbig Ae/Be stars (intermediate mass young
stars) which can be compared to our T Tauri sample.  For the Herbig Ae
and late Be objects, \citet{mil01} and \citet{eis03} found
near-infrared disk sizes consistent with an inner disk edge at the
dust sublimation radius set by direct exposure to radiation from the
central star; while the early Be objects have near-infrared disk sizes
which are relatively smaller, implying that some material is present
interior to the dust sublimation radius \citep{mon02,eis04,mon05}.
The \citet{mon05} Herbig study used the \citet{mon02} dust sublimation
(referred to here as the optically thin model, f=0.047 in
Eq. \ref{Rdust}) and found sizes consistent with dust sublimation
temperatures of 1000 to 1500~K, in general agreement with our T Tauri
results.  Accretion luminosities were not included in the Herbig
comparison, but the properties of Herbig stars can be modeled with
low accretion rates \citep[$\dot{M}=10^{-7}$ M$_{\odot}$/yr; e.g.][]
{muz04} which would result in very low accretion
luminosities compared to their stellar luminosities.

Although the sample in this paper is small with substantial
uncertainties in the individual sources, a range of inner dust disk
properties is suggested by these results.  In the right panel of
Figure \ref{fig:dust} we explore the discrepancy of the models by
plotting R$_{measured}$/R$_{dust}$ vs. L$_{accretion}$/L$_{star}$,
where R$_{dust}$ is the mean of the two models in the left panel and
the error bar is the range between the optically thin and thick
models.  The sources with low accretion to stellar ratios generally
have radius ratios larger than 1 (GM Aur has the highest radius ratio)
while the sources with roughly equal accretion and stellar
luminosities have radius ratios near 1.  If the accretion rate is a
proxy for age in young stellar systems \citep[see e.g.][]{har98}, this
range might be evolutionary in nature.  For example, work by
\citet{cla01} modeling the interaction of photoevaporation and
accretion has shown that the inner disk can dissipate before the outer
disk.  Observations of more sources are needed to further explore this
possibility.

\acknowledgments

The Keck Interferometer is funded by the National Aeronautics and
Space Administration as part of its Navigator program.  Part of this
work was performed at the Michelson Science Center and the Jet
Propulsion Laboratory, California Institute of Technology, under
contract with NASA.  Observations presented were obtained at the
W.M. Keck Observatory, which is operated as a scientific partnership
among the California Institute of Technology, the University of
California and NASA.  The Observatory was made possible by the
generous financial support of the W.M. Keck Foundation.  The authors
wish to recognize and acknowledge the very significant cultural role
and reverence that the summit of Mauna Kea has always had within the
indigenous Hawaiian community.  We are most fortunate to have the
opportunity to conduct observations from this mountain.  We thank
Aletta Tibbetts for help in acquiring the MDM data.  This material is
based upon work supported by NASA under JPL Contracts 1236050 \&
1248252 issued through the Office of Space Science.  This work has
made use of software produced by the Michelson Science Center at the
California Institute of Technology, the SIMBAD database, operated at
CDS, Strasbourg, France, and the NASA/IPAC Infrared Science Archive,
operated by the JPL under contract with NASA.  This publication makes
use of data products from the Two Micron All Sky Survey, which is a
joint project of the University of Massachusetts and the Infrared
Processing and Analysis Center/California Institute of Technology,
funded by the National Aeronautics and Space Administration and the
National Science Foundation.

\end{document}